\documentclass[useAMS,usenatbib]{mn2e}

\usepackage{graphicx}

\defcitealias{farihi13}{FGK13}

\title[Radiative Levitation of Silicon]{Radiative Levitation of Silicon in the Atmospheres of Two Hyades DA White Dwarfs}

\author[P. Chayer]{P. Chayer\thanks{E-mail:
chayer@stsci.edu}\\
Space Telescope Science Institute, 3700 San Martin Drive, Baltimore, MD 21218, USA}

\begin{document}



\maketitle

\label{firstpage}

\begin{abstract}
The presence of silicon at the surface of the two Hyades DA white dwarfs WD~0421+162 and WD~0431+126 requires mechanisms that counteract the effects of the downward diffusion. Radiative levitation calculations indicate that the silicon abundance observed in WD~0431+126 corresponds to the abundance supported by radiative levitation. Detailed time-dependent diffusion calculations that take into account radiative levitation and accretion indicate that accretion with rates of $\dot{M}_{\rm{Si}} \la 10^4$~g~s$^{-1}$ could also be present without disrupting the abundance supported by radiative levitation. In the case of WD~0421+162, accretion with a rate of $\dot{M}_{\rm{Si}} = 10^{5.1}$~g~s$^{-1}$ must be invoked, because its observed silicon abundance is larger than the abundance supported by radiative levitation. This accretion rate is lower than the accretion rate given by the accretion-diffusion model, because the radiative levitation slows down the downward diffusion of silicon. The silicon abundances observed in the two Hyades white dwarfs cannot be interpreted solely in terms of accretion. The interpretation of the silicon abundances must take into account the interplay between radiative levitation and accretion. 
\end{abstract}

\begin{keywords}
white dwarfs: individual(WD~0421+162, WD~0431+126) --- stars: chemically peculiar --- diffusion --- accretion.
\end{keywords}

\section{Introduction}
The observations of metals in the atmospheres of white dwarfs offer a unique opportunity to explore the nature of the circumstellar matter that surrounds white dwarfs. Because of the overwhelming efficiency of the gravitational settling, most DA white dwarfs have pure-hydrogen atmospheres. When traces of metals are observed, a physical mechanism must counteract the effects of the gravitational settling to maintain the metals in the photosphere, otherwise they diffuse rapidly out of the photosphere. Accretion of circumstellar matter has been invoked in several instances to explain traces of metals that are observed in the photospheres of white dwarfs \citep*[see, e.g.,][]{becklin_etal05, gansicke_etal06, vonhippel_etal07, farihi_etal08, vennes_etal10, dufour_etal12}. In the presence of accretion, material falls onto the surface of a white dwarf and diffuses downward until a steady state is reached.  An abundance is then maintained in the photosphere of the star as the accretion replenishes the matter that diffuses downward. The accretion-diffusion model relates the photospheric abundances to the abundances in the accreted matter  \citep[see, e.g.,][]{dupuis_etal1993,koester_wilken2006}. Therefore, by analysing  abundances in a white dwarf that is accreting, we can characterize not only the properties of the accreted matter, but also deduce the nature of the circumstellar matter.

\citet*{farihi13}, hereafter FGK13, took full advantage of the accretion-diffusion model to suggest that rocky planetesimals and small planets had once orbited two main-sequence A-type stars in the Hyades open cluster. \citetalias{farihi13} based their findings on observations of two Hyades DA white dwarfs that were obtained with the Cosmic Origins Spectrograph (COS) onboard the {\it Hubble Space Telescope}. They analysed the COS spectra of the DA white dwarfs WD~0421+162 and WD~0431+126, and measured silicon abundances and stringent carbon abundance upper limits. By assuming that both stars are currently in an accretion-diffusion equilibrium, they concluded that the accreted matter was more carbon deficient than chondritic meteorites.  They suggested that the circumstellar matter that is being accreted could have a rocky composition. 

The main drawback of the \citetalias{farihi13} analysis is their dismissal of the radiative levitation of silicon. \citet{chayer_dupuis2010} and \citet*{dupuis_etal2010} suggested that the radiative levitation of silicon in the atmospheres of DA white dwarfs with 17,000~K $\la T_{\rm{eff}} \la 25$,000~K could affect the determination of accretion rates in white dwarfs when one uses the accretion-diffusion model.  By carrying out time-dependent diffusion calculations that included radiative levitation and accretion, \citet{chayer_dupuis2010} demonstrated that the silicon abundance observed in the atmosphere of a white dwarf with $T_{\rm{eff}} = 20$,000~K and $\log g = 8.0$ is affected by radiative levitation. Given that these atmospheric parameters are close to those of WD~0421+162 and WD~0431+126, this suggests that the radiative support of silicon in both Hyades stars could be important.

The purpose of this paper is to extend the calculations of \citet{chayer_dupuis2010} to WD~0421+162 and WD~0431+126, and to show that radiative levitation must be included in the interpretation of the silicon abundances. I first show in \S{\ref{levitation}} that the radiative support of silicon is not negligible in the atmospheres of the two stars. By using this result, I present in \S{\ref{diff_calculations}} the results of time-dependent diffusion calculations that take into account radiative levitation and accretion. In \S{\ref{discussion}}, I discuss the problems that \citetalias{farihi13} put forward on the interpretation of the radiative support of silicon in the Hyades DA white dwarfs,  and conclude in \S{\ref{conclusion}}. 

\section{Diffusion of Silicon}\label{silicon}

\subsection[]{Radiative Levitation: Equilibrium Calculations}\label{levitation}

Several authors demonstrated that radiative levitation is an important ingredient when trying to explain the appearance of metals in the atmospheres of hot white dwarfs \citep*[see, e.g.,][]{vvg,cfw95,dreizler_wolff99}.  In particular, the calculations of \citet{cfw95} and \citet{chayer_dupuis2010} showed that traces of silicon and aluminum could be maintained in the atmospheres of DA white dwarfs as cool as $T_{\rm{eff}} = 20$,000~K. Based on the same theoretical framework that was developed by \citet{cfw95}, I computed radiative acceleration of silicon in DA models corresponding to WD~0421+162 and WD~0431+126. The atmospheric parameters that were used correspond to those that \citetalias{farihi13} have adopted: $T_{\rm{eff}} = 19$,242~K and  $\log g = 8.09$ for WD~0421+162; $T_{\rm{eff}} = 21$,202~K and $\log g = 8.11$ for WD~0431+126. 

\begin{figure}
\includegraphics[width=87mm]{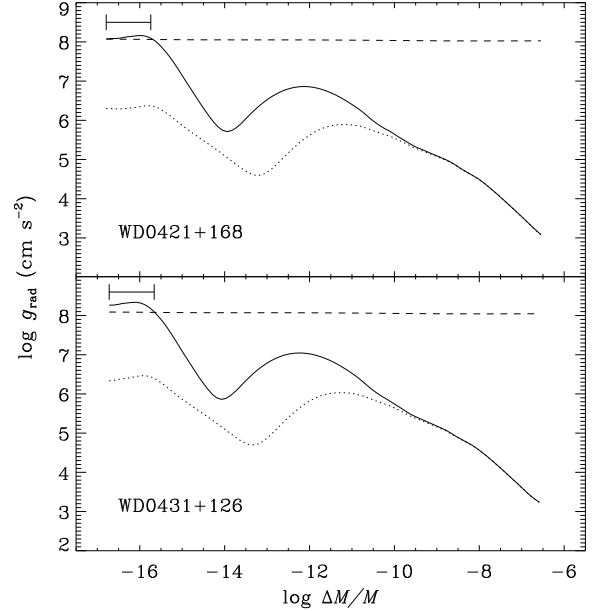}
\caption{Radiative acceleration of silicon as a function of depth in DA models corresponding to WD~0421+162 and WD~0431+126. The solid (dotted) curve is the radiative acceleration of silicon that is computed for a mass fraction $X_{\rm{Si}} = 10^{-8}$ ($X_{\rm{Si}} = 10^{-2}$). The solid curve is the maximum radiative acceleration of silicon that illustrates  the case for which the silicon lines are completely desaturated. On the other hand, the dotted curve illustrates a case when the silicon lines are saturated. The dashed line is the local effective gravity. The horizontal solid line indicates the region where silicon can be supported by radiative levitation.
\label{sigrad}}
\end{figure}

Figure~\ref{sigrad} illustrates radiative acceleration ($g_{\rm{rad}}$) of silicon as a function of depth in models of WD~0421+162 and WD~0431+126. The solid and dotted curves are radiative accelerations computed for two silicon mass fractions of $X_{\rm{Si}} = 10^{-8}$ and $10^{-2}$. The radiative acceleration at the top of the envelopes is computed at an optical depth of $\tau = \frac{2}{3}$. The horizontal dashed line is the effective local gravity (see, equation(\ref{eq_geff})). The radiative acceleration computed for a mass fraction of $X_{\rm{Si}} = 10^{-8}$ is the theoretical maximum, and corresponds to the regime when the silicon lines are completely desaturated, so that the silicon atoms can absorb the maximum momentum coming from the radiation field. Figure~\ref{sigrad} shows that for both stars the radiative acceleration can be greater than or equal to the gravitational acceleration for depths of $\log \Delta M/M \la -15.5$. A reservoir of silicon at the surface of  WD~0421+162 and WD~0431+126 can therefore be supported by radiative levitation. 

The expected surface abundances of silicon in both DA white dwarfs were computed by estimating the abundances for which the radiative acceleration is equal to the effective gravity, which is given by 

\begin{equation}
g_{\rm{eff}} = \left[1-\frac{A_{\rm{H}}(Z_{\rm{Si}}+1)}{A_{\rm{Si}}(Z_{\rm{H}}+1)}\right]g,\label{eq_geff}
\end{equation}

\noindent where $A_{\rm{H}}$ and $A_{\rm{Si}}$ are the atomic weight of hydrogen and silicon, $Z_{\rm{H}}$ and $Z_{\rm{Si}}$ are the average charges of hydrogen and silicon, and $g$ is the local gravity. The expected silicon abundances were computed at $\tau = \frac{2}{3}$ given that the abundances must be compared to the observed silicon abundances. I determined silicon abundances of 
$\log N({\rm{Si}})/N({\rm{H}}) = -9.8$ and $-8.1$ in the photospheres of WD~0421+162 and WD~0431+126.  These abundances can be compared with $-7.5$ and $-8.0$, respectively determined by \citetalias{farihi13} for the same stars. The expected silicon abundance in WD~0431+126 agrees almost perfectly with the observed abundance, indicating that radiative levitation alone could explain the appearance of silicon. On the other hand, the silicon abundance observed in WD~0421+162 is much larger than that predicted by the radiative levitation theory. This fact suggests that a second mechanism must be at work to supplement the additional amount of silicon observed in WD~0421+162. \citet{cfw95, chayer_etal95} have proposed that weak stellar winds could disrupt the radiative equilibrium abundances in hot DA white dwarfs with $T_{\rm{eff}} \ga 40$,000~K, and could explain the discrepancies between the observed and predicted abundances. \citet{rauch_etal2013} have reached a similar conclusion when analysing the photospheric abundances of G191-B2B ($T_{\rm{eff}} = 60$,000 K) with their self-consistent diffusion model. However, we do not expect the presence of weak stellar winds in cooler stars such as WD~0421+162. Therefore, the most plausible mechanism that could provide additional Si in the atmosphere of WD~0421+162 is accretion.

\subsection{Radiative Levitation and Accretion: Time-Dependent Calculations}\label{diff_calculations}

When accretion and radiative levitation are at work in the atmosphere of a white dwarf, the observed silicon abundance cannot be easily related to the silicon abundance in the accreted matter. In order to estimate the effects of both accretion and radiative levitation on the photospheric abundance, one must carry out time-dependent diffusion calculations that follow the changes in the silicon abundance as functions of radius and time. The physical problem is described by an initial-boundary-value diffusion equation, where accretion is considered as a source term at the surface of the star, and where radiative acceleration is considered as a term in the microscopic diffusion velocity. The accretion is assumed to be spherically symmetric. By solving this diffusion equation one can follow the silicon abundance in the envelope of a white dwarf as a function of time. Similar calculations have been carried out in white dwarf models by \citet{dupuis_etal1993} who considered diffusion in the presence of accretion, by \citet{vennes_etal88} who considered diffusion of helium in the presence of radiative levitation, and by \citet*{chayer_etal97} who considered diffusion of silicon in the presence of radiative levitation and mass loss. \citet{chayer_dupuis2010} were the first to carry out time-dependent diffusion calculations in DA white dwarfs that included accretion and radiative levitation.

\begin{figure}
\includegraphics[width=87mm]{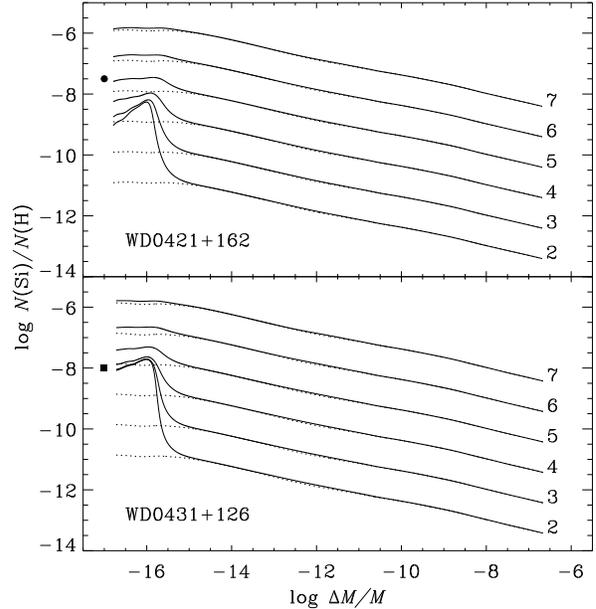}
\caption{Steady state distributions of silicon in stellar models corresponding to WD~0421+162 and WD~0431+126 in the presence of continuous accretion at rates of $\dot{M}_{\rm{Si}} = 10^2$, $10^3$, $10^4$, $10^5$, $10^6$, and $10^7$ g s$^{-1}$. The curves are labeled by the logarithmic value of $\dot{M}_{\rm{Si}}$. The calculations are carried out by considering accretion in the presence of radiative levitation ({\it solid curve}), and by considering only accretion ({\it dotted curve}). The filled circle and square symbols indicate the silicon abundances observed by \citetalias{farihi13}. These abundances must be compared to the predicted abundances at the top of the envelopes.  
\label{wd_comp_acc}}
\end{figure}

All the present diffusion calculations were carried out in stellar envelope models corresponding to the \citetalias{farihi13} parameters of WD~0421+162 and WD~0431+126. An initial homogeneous silicon abundance of $\log N({\rm{Si}})/N({\rm{H}}) = -7.0$ was considered throughout the stellar envelopes. The calculations were carried out for several accretion rates ranging from $10^2$ to $10^7$ g s$^{-1}$. The top of the envelope corresponds to $\tau = \frac{2}{3}$. Because accretion is treated as a source term at the surface of the star, silicon is assumed to be uniformly mixed in the optically thin layers above the optical depth $\tau = \frac{2}{3}$. Radiative accelerations were computed according to \citet{cfw95}. The time-dependent diffusion calculations were performed in order to cover a period of time $t = 10^6$ yr. This period of time is long enough for attaining a steady-state distribution of silicon in layers above $\log \Delta M/M \la -5.0$. Figure~\ref{wd_comp_acc} illustrates the main results. The figure shows steady-state distributions of silicon in stellar models of WD~0421+162 and WD~0431+126 in the presence of accretion for six accretion rates.  Each curve represents the steady-state distribution of silicon for a given accretion rate. The dotted curves are steady-state distributions of silicon in the presence of accretion but without radiative levitation. The solid curves are steady-state distributions of silicon in the presence of accretion and radiative levitation. 

The dotted curves in Figure~\ref{wd_comp_acc} show that silicon can be maintained by accretion in the upper layers of WD~0421+162 and WD~0431+126. A steady-state is reached when the number of silicon atoms that are accreted is equal to the number of atoms that diffuse downward. A silicon abundance equilibrium is reached at each depth even though the atoms continue to diffuse downward. This abundance equilibrium is possible, because silicon atoms are continuously replenished by accretion as they diffuse downward. The dotted curves show that the abundances depend on the accretion rates, so a measure of the abundance at the surface of a white dwarf can be related directly to the accretion rate. By performing a linear interpolation on the accretion rates at the abundances observed in WD~0421+162 and WD~0431+126, I estimated accretion rates of $\log \dot{M}_{\rm{Si}} = 5.4$ and 4.9, respectively. These accretion rates are essentially equal to those of \citetalias{farihi13} determined by assuming the accretion-diffusion model. 

In the presence of radiative levitation, however, the observed silicon abundances cannot be simply related to the accretion rates. This is illustrated in Figures~\ref{wd_comp_acc} by the solid curves that give the steady-state abundances computed in the presence of accretion and radiative levitation.  The difference between the dotted and solid curves is obvious above the depths $\log \Delta M/M \la -15.5$, where the radiative support is important.  The main difference is the appearance of a silicon reservoir that results from the equilibrium between the upward radiative acceleration and the downward gravitational acceleration of silicon. At low accretion rates, the equilibrium abundances are essentially given by the radiative levitation alone. In particular, the lower panel of Figure~\ref{wd_comp_acc} shows that the surface abundance predicted by the radiative levitation matches the silicon abundance observed in the atmosphere of WD~0431+126. The abundance predicted by the radiative levitation holds on even in the presence of accretion with rates up to  $\log \dot{M}_{\rm{Si}} \simeq 4.0$ (g~s$^{-1}$). 

The upper panel of Figure~\ref{wd_comp_acc} shows, on the other hand, that the radiative levitation alone cannot explain the observed silicon abundance in WD~0421+162. The figure shows that the radiative levitation can support an abundance close to $\log N({\rm{Si}})/N({\rm{H}}) \simeq -9.1$, which is much lower than the observed abundance. The figure indicates that an accretion rate of $\log \dot{M}_{\rm{Si}} = 5.1$ is necessary to maintain the observed silicon abundance. It is noteworthy that this accretion rate is smaller than the accretion rate ($\log \dot{M}_{\rm{Si}} = 5.4$) needed when the radiative levitation is neglected. This phenomenon is due to the effects of the radiative levitation on the downward diffusion of silicon. As silicon is accreted and diffuses downward, the radiative levitation slows down its downward diffusion and helps to maintain a larger amount of silicon in the upper layers of the star. 

\section{Discussion}\label{discussion}

After having demonstrated that radiative levitation has an impact on the calculations of silicon accretion rates in stars such as WD~0421+162 and WD~0431+126, we can address the issues that \citetalias{farihi13} raised with the interpretation of the radiative levitation.  We will review their arguments and see whether they are appropriate in the light of the current and past calculations. Their first argument against radiative levitation was the fact that the cooler of the two stars had a higher silicon abundance, in contrast to the prediction of the levitation theory. The upper panel of Figure~\ref{wd_comp_acc} shows that radiative levitation alone cannot support the whole amount of silicon that is maintained in the atmosphere of WD~0421+162. Accretion must be at work to supply the additional silicon. This instance demonstrates that both accretion and radiative levitation are at work simultaneously to maintain silicon in the atmosphere of the star. In this case, even though accretion is the dominant mechanism to maintain silicon, radiative levitation nevertheless contributes to the support of silicon. 

The second problem that \citetalias{farihi13} raised concerns the low silicon abundances that are observed in seven stars, which are part of their COS Snapshot survey \citep[G\"ansicke et al. 2013, in preparation;][]{koester_etal2013}. According to \citetalias{farihi13}, these stars have silicon abundances $\log N({\rm{Si}})/N({\rm{H}}) < -8.0$ and should have a greater radiative support than the one predicted in the stars WD~0421+162 and WD~0431+126. It is difficult to comment on these stars, because \citetalias{farihi13} did not give any specific informations on the uncertainties of the atmospheric parameters. \citet{koester_etal2013} report the silicon abundances, but do not provide any information on the stars' atmospheric parameters. What I can say, though, is that the silicon abundances predicted by the radiative levitation theory are sensitive to temperature and gravity, especially in the temperature range of the two Hyades white dwarfs. In this temperature range, the predicted silicon abundances decrease rapidly as the temperature decreases. A slight change in temperature or gravity could change significantly the predicted silicon abundances. 
Therefore, detailed radiative levitation calculations that include uncertainties on the stellar atmospheric parameters could shed light on this issue. 

The observations of the seven stars for which the abundances are lower than the predicted abundances could also indicate that other phenomena are at work at the surface of the stars. For example, the assumption that accretion is spherically symmetric could be wrong. We could imagine matter that is accreted and forced to follow a star's magnetic field. This matter would then pile up at its magnetic poles. In this case, measured abundances would be underestimated and would not match predicted abundances.   Another plausible phenomenon that could have an effect on predicted abundances is turbulence. Although turbulence is not well understood at the surface of white dwarfs, its effect would disrupt radiative levitation, and consequently, affect predicted abundances.   

\citetalias{farihi13}'s third argument concerns the presence of silicon during the earlier phases of the two Hyades white dwarfs. \citetalias{farihi13} argue that silicon was not supported by radiative levitation during the whole cooling period of the stars, and therefore it must be external. The calculations carried out in \S\ref{diff_calculations} demonstrate that silicon does not need to be primordial in order to be supported by radiative levitation. The lower panel of Figure~\ref{wd_comp_acc} shows that very low accretion rates can provide the necessary silicon. For low accretion rates, the predicted abundance in a star such as WD~0431+126 will be the abundance maintained by radiative levitation. Figures~\ref{sigrad} and \ref{wd_comp_acc} show that whether silicon is external or primordial, once it is in the atmospheres of the two white dwarfs, it will definitely be subject to radiative levitation. 
 
Finally, \citetalias{farihi13} mention that previous results from radiative levitation have not yet been quantitatively successful for white dwarfs with $T_{\rm{eff}} > 25$,000~K. Indeed, \citet{cfw95, chayer_etal95} concluded that the equilibrium radiative levitation theory fails to account for the observed abundances. This conclusion lead \citet{cfw95,chayer_etal95} to point out, however, that other mechanisms must be simultaneously at work in these stars. They suggested mechanisms such as weak stellar winds, turbulence, and accretion that could compete with gravitational settling and radiative levitation. They emphasized that in spite of the failure of the equilibrium radiative levitation theory at explaining the observed abundances, radiative levitation cannot be ignored in hot white dwarfs and must be included as a basic ingredient in more elaborated models. As we have seen in \S{\ref{silicon}}, this fact applies to silicon in WD~0421+162 and WD~0431+126.

\section{Conclusion}\label{conclusion}

According to models presented here, the silicon abundances observed in the atmospheres of the DA white dwarfs WD~0421+162 and WD~0431+126 cannot be directly related to the composition of the accreted matter, because radiative levitation plays an important role in supporting silicon. The amount of silicon observed in the atmosphere of WD~0431+126 could be entirely supported by radiative levitation, although a small amount of silicon could be accreted. In the star WD~0421+162, the observed silicon abundance is significantly higher than the abundance predicted by the radiative levitation, so accretion of silicon must be invoked. As silicon is accreted, the radiative levitation slows down its downward diffusion, and reduces the amount of necessary external silicon. The expected accretion rate is a factor of 2 lower than the rate that is given by the accretion-diffusion scenario. Accretion rate calculations carried out without radiative levitation can have a significant impact on the interpretation of the chemical composition of the accreted matter, and consequently, on the interpretation of its origin.

\section*{Acknowledgments}

I wish to thank Gilles Fontaine and Jean Dupuis for their useful comments. I acknowledge financial support by the Canadian Space Agency under a contract with the National Research Council Canada, National Science Infrastructure in Victoria, British Columbia, Canada.


\bsp

\label{lastpage}

\end{document}